# Selective etching of (111)B-oriented $Al_xGa_{1-x}As$-layers for epitaxial lift-off

*Tobias Henksmeier\*, Martin Eppinger, Bernhard Reineke, Thomas Zentgraf, Cedrik Meier and Dirk Reuter*

T. Henksmeier, M. Eppinger, B. Reineke, Prof. Dr. T. Zentgraf, Prof. Dr. C. Meier, Prof. Dr. D. Reuter
Paderborn University, Department of Physics, Warburger Str. 100, 33098 Paderborn, Germany
E-mail: tobias.henksmeier@mail.upb.de

**GaAs-(111)-nanostructures exhibiting second harmonic generation are new building blocks in nonlinear optics. Such structures can be fabricated through epitaxial lift-off employing selective etching of Al-containing layers and subsequent transfer to glass substrates. In this article, the selective etching of (111)B-oriented $Al_xGa_{1-x}As$ sacrificial layers (10 nm to 50 nm thick) with different aluminum concentrations (x=0.5 to 1.0) in 10 % hydrofluoric acid is investigated and compared to standard (100)-oriented structures. The thinner the sacrificial layer and the lower the aluminum content, the lower the lateral etch rate. For both orientations, the lateral etch rates are in the same order of magnitude, but some quantitative differences exist. Furthermore, the epitaxial lift-off, the transfer, and the nano-patterning of thin (111)B-oriented GaAs membranes is demonstrated. Atomic force microscopy and high-resolution x-ray diffraction measurements reveal the high structural quality of the transferred GaAs-(111) films.**

## 1. Introduction

Recently, second harmonic generation (SHG) in nanoparticles and nanostructures has gained much interest as a platform for nonlinear optics.[1-5] GaAs-nanostructures with (111)-surface orientation are attractive candidates as they exhibit efficient forward directed emission while for (100)-oriented GaAs, there is a strong pump pulse polarization dependence hindering efficient SHG emission perpendicular to the surface.[1,6,7] Furthermore, a high refractive index contrast enhances SHG from GaAs nanostructures.[1,5] Therefore, the nanostructures must be prepared on a transparent substrate with a low refractive index [1,5] to provide high optical mode confinement towards the substrate. One route to achieve this was demonstrated by Sautter and co-workers, who performed epitaxial lift-off of an array of nanostructures and bonded them to a suitable substrate.[1] Another fabrication route, which was demonstrated by Person and co-workers for (100)-oriented GaAs structures [8] and which we will discuss in this publication for (111)-oriented GaAs films, is the transfer of a thin membrane by a combination of epitaxial lift-off and bonding followed by subsequent nano-patterning.

Epitaxial lift-off of GaAs-based heterostructures was first investigated by Yablonovitch and co-workers, who transferred GaAs-based heterostructures in (100)-orientation.[9,10] Black wax was applied onto small samples before a thin buried $Al_xGa_{1-x}As$ sacrificial layer with high aluminum-content was etched by hydrofluoric acid (HF). The so released films were then transferred and bonded to various substrates.[10] Until now, numerous investigations were made on the selective etching of $Al_xGa_{1-x}As$ sacrificial layers with (100)-orientation, studying e.g. different HF-concentrations and temperatures, aluminum concentrations in the sacrificial layer, or the reaction kinetics in the lateral etch channel.[11-14] Besides, sophisticated methods to lift-off complete wafers were reported.[15, 16] Practically all studies so far have been done for the technologically relevant (100) surface orientation. In one case, epitaxial lift-off has been demonstrated for (111)-oriented layers.[1] However, the focus of that work was on the optical properties of these nanostructures and the epitaxial lift-off has not been investigated systematically. In this article, we present a study on the lateral etch rates of $Al_xGa_{1-x}As$ sacrificial layers with (111)B-orientation for different layer thicknesses (10 nm to 50 nm) and aluminum contents (x=0.5 to 1.0). The lift-off and transfer of 500 nm thick GaAs-(111)B films to glass substrates and subsequent nano-patterning are demonstrated, where the high structural quality of the film is maintained after the transfer.

## 2. Experimental Details

All samples were grown by solid-source molecular beam epitaxy on GaAs-(111)B substrates with a 1 ° miscut towards (211). Besides, samples on GaAs-(100) substrates with nominally on-axis orientation have been fabricated for comparison. The layer sequence starts with a 100 nm GaAs buffer layer followed by an $Al_xGa_{1-x}As$ sacrificial layer of variable thickness d (10 nm ≤ d ≤ 50 nm) and is terminated by 500 nm GaAs. The 500 nm GaAs is the layer we will lift-off. The aluminum-content x of the sacrificial layer was varied between 50 % and 100 %. As already known from the literature [17-20], the process window for the growth of GaAs and even more for $Al_xGa_{1-x}As$ on GaAs(111)B-substrates is rather small to obtain smooth surfaces. We employed a growth temperature of 590 °C and beam equivalent arsenic (As$_4$) pressure of $p_{As4} = 1.50 \times 10^{-5}$ mbar at a growth rate for GaAs of 0.3 monolayers per second (ML/s). To obtain smooth AlAs films, the As$_4$-pressure had to be decreased compared to the growth of pure GaAs to $p_{As4}=0.90 \times 10^{-5}$ mbar and the substrate temperature had to be increased to 630 °C. The growth rate was 0.15 ML/s. For smooth $Al_xGa_{1-x}As$ layer growth, the arsenic pressure and the growth temperature was linearly interpolated between the arsenic pressure and the growth temperature used for GaAs and AlAs, respectively. For $Al_xGa_{1-x}As$, the growth rate varied from 0.15 ML/s to 0.45 ML/s, depending on the Al-content x. With optimized growth parameters, a smooth surface with a surface roughness (root-mean-square roughness) of $\leq 0.5$ nm was obtained on a $1\times1$ µm$^2$ area and $\leq 0.7$ nm for $10\times10$ µm$^2$ (see **Figure 1**a for an example). The roughness values are average values for AFM measurements taken at different positions of the sample. For the growth of (100)-oriented substrates, the growth window is wider and we employed a substrate temperature of 600 °C and an $As_4$-pressure of $p_{As4} = 2.27\times10^{-5}$ mbar for the growth of the whole layer sequence. The growth rate for GaAs was 0.7 ML/s and for $Al_xGa_{1-x}As$ between 0.35 ML/s and 1.05 ML/s.

After growth, small samples of $4\times4$ mm$^2$ edge length were cleaved from the wafers. Due to the different orientations of the easy cleaving directions, for the (100)-orientation we obtained square-shaped samples whereas for the (111)-orientation diamond-shaped samples with a 60 ° angle were prepared.

The samples were processed in two ways:

(1) **Figure 2** shows the process schematically. Black wax is applied to the sample's surface following a procedure developed by Yablonovitch and co-workers.[9,10] Then, the samples were immersed into 10 % HF at room temperature. Due to the high selectivity of hydrofluoric acid, only the $Al_xGa_{1-x}As$ sacrificial layer is laterally etched. The GaAs is practically not etched at all. As the wax applies stress to the sample, the under-etched GaAs layer bends upwards during $Al_xGa_{1-x}As$ sacrificial layer etching, as shown in Figure 2, and the etch channel is opened. The reaction product diffusion is enhanced, and etch stopping due to a clogged channel is hindered. After various time intervals of 1 h – 4 h, etching was stopped by dipping the samples into deionized water (DI-water). With this etching duration, the sacrificial layer was not completely removed but in the center of the sample, the $Al_xGa_{1-x}As$-layer remained in a certain area (see Figure 2b). We removed the wax by toluene employing an ultrasonic bath for several minutes. As the stabilizing wax is removed, the fragile under etched GaAs film is torn off and only in the area still connected by the $Al_xGa_{1-x}As$-layer, the GaAs remained (see Figure 2c). As a consequence, a clear step separating the under-etched from the not under-etched region becomes visible as exemplarily shown in **Figure 3**. The lateral etch depth $d_{lateral}$ can then be determined under an optical microscope measuring the distance between the sample edge and the step on the surface ($d_{lateral}$ is indicated in Figures 2b and 2c as well as in Figure 3 by a green arrow). In the experiment, the step is not straight over the sample length, as can be seen in Figure 3. For the evaluation of $d_{lateral}$, we approximated the step over the whole sample length by a straight line parallel to the sample edge at an intermediate position. Employing this straight line, the average distance to the sample edge was determined and an error for this value of $d_{lateral}$ was estimated. The lateral etch rate is then calculated from the experimentally determined average lateral etch depth $d_{lateral}$ and the etching time the sample was immersed in the HF solution. The calculated lateral etch rate is an average lateral etch rate for the etching time employed and decreases with increased etching time. The error in the lateral etch rates is approximately ± 20 %, which is quite large. However, this was also observed in the literature for (100)-oriented samples. [13,21] We etched all sacrificial layer types for two different times (65 and 130 minutes for most Al-concentrations, only for $x \leq 60$ % we etched up to 260 minutes) and determined the individual average lateral etch rates. Another option to determine the lateral etch rate is to measure the time for a complete lift-off of the GaAs film in the hydrofluoric acid. However, we observed that the completely under-etched GaAs layers stick weakly to the substrates and do not float on the etchant surface immediately after complete removal of the sacrificial layer. This results in a significant scattering of the values for the etching times and in conclusion of the lateral etch rates, so this method was not used for the quantitative values presented here. To test the transfer of the 500 nm thick GaAs membrane, we removed the sacrificial layer entirely, and the floating film was carefully transferred to a glass substrate and dried in a desiccator for roughly 24 h. Then the wax was removed in toluene (without employing an ultrasonic bath). The bonded samples were then baked in a vacuum oven (p ≈ 5 mbar) for roughly 24 h at 140 °C to strengthen the bonds between film and glass. To remove wax residuals, we applied an oxygen ashing process for 90 seconds followed by a short HCl dip to remove surface oxides. AFM (see **Figure 1**b) and HRXRD measurements revel the intactness of the bonded film. Finally, for a proof-of-principle experiment, GaAs-pillars on glass were fabricated using standard electron beam lithography and reactive ion etching employing a SiN hard mask.

(2) We have also used a method following the approach by Voncken et co-workers [12], which in principle is less time-consuming. Mesa-structures were prepared by optical lithography and reactive ion etching (etch depth approximately 2 μm to expose the $Al_xGa_{1-x}As$ sacrificial layer at the sidewalls of the mesa). By this preparation, the $Al_xGa_{1-x}As$-layer are exposed at the sidewalls. In the next step, the samples were etched for a fixed time in 10 % HF. After etching, the samples were cleaved through the mesa-structure and the cleavage plane was analyzed by scanning electron microscopy (SEM). The lateral etch depths $d_{lateral}$ were determined from the SEM images (not shown here) and the lateral etch rates were calculated by dividing the lateral etch depth $d_{lateral}$ by the etch time. The values for the lateral etch rates determined with this method showed a significantly larger scattering compared to method (1), so no quantitative values are reported here. However, the results follow the same overall trend as those obtained with method (1) and support the quantitative findings presented in the following. We observe further that without supporting wax, the GaAs layers sometimes stick to the substrates and that the etch channel is clogged for longer etch times.

## 3. Results and discussion

**Figures 4a** and **4b** summarize the lateral etch rates of the (111)-oriented $Al_xGa_{1-x}As$ sacrificial layers ($0.5 \leq x \leq 1$) in 10 % HF determined according to method (1) for two different sacrificial layer thicknesses of 25 nm and 50 nm. For comparison, **Figure 4c and 4d** show the results obtained for (100)-oriented samples. A general trend is observed for both orientations: The higher the aluminum concentration, the higher the lateral etch rate. No etching is observed for aluminum concentrations $\leq 50$ % for both orientations. Whereas for 60 % aluminum content etching starts already for (111)-oriented sacrificial layers (at least for 50 nm thickness), etching of the (100)-oriented sacrificial layers is not observed for this concentration. For 70 % aluminum content, significant etching is observed for both orientations. This means that we should be able to lift-off Al-containing heterostructures with up to 50 % Al-content for both orientations with the employed process parameters. It seems, for (100)-oriented layers even heterostructures with slightly higher Al-concentration can be selectively under-etched.

A rapid increase of the lateral etch rate for both, the 25 nm thick and the 50 nm thick (111)-oriented sacrificial layers is observed for aluminum concentrations $\geq 70$ %, e.g., the lateral etch rate of a 50 nm thick $Al_{70}Ga_{30}As$-(111) sacrificial layer is already $(510\pm100)$ µm h$^{-1}$ and sufficient to obtain complete layer lift-off within several hours. In comparison, the lateral etch rates of the 50 nm thick (100) oriented sample is roughly a factor 3 – 4 smaller. At higher aluminum concentrations $\geq 85$ % the lateral etch rates for the (111)-oriented sacrificial layers saturate while they still increase with Al-content for the (100)-oriented layers. The maximum lateral etch rate for the (111)-oriented layers is approximately $(625\pm100)$ µm h$^{-1}$ and is already reached for 85 % Al-content. For the (100)-oriented reference samples, the lateral etch rates increases from 85 % to 100 % Al-content by approximately 20 % – 30 %. The maximum lateral etch rate of $(640\pm100)$ µm h$^{-1}$ is reached for 100 % aluminum.

In summary, under our experimental conditions, the maximum lateral etch rates are the same for both orientations within the error bars and sufficiently high that epitaxial lift-off on a mm-scale is feasible in reasonable times. However, for aluminum concentrations < 100 %, the lateral etch rates drop faster for the (100)-oriented sacrificial layers than for the (111)-orientation. This allows (111)-oriented sacrificial layers with lower Al-content, but also provides slightly less selectivity for the lift-off of heterostructures with high Al-content. The reason for the observed differences might be due to the different arrangements on the atomic scale. This might either affect the chemical reaction due to the different bond orientations and atom density or the response to the applied stress by the wax due to different relevant elasticity constants. Different response to stress would lead to differences in the etch channel geometry, i.e., the width of the channel as a function of its depth. This might also be caused by the different shapes of the sample (squares vs. diamonds) so that the stress due to the wax might be different for the two sample orientations.

For different etch times, we observe that the average lateral etch rates tend to decrease for longer etching times independent of sample orientation. For the 50 nm thick $Al_xGa_{1-x}As$-(111) sacrificial layers with aluminum contents ≥ 70 %, doubling the etch time results in a lateral etch rate decrease by roughly 20 % - 25 % independent of the aluminum concentration (see Figure 4) while there is no change for 25 nm thick sacrificial layers. In comparison, for the 25 nm and 50 nm (100)-oriented sacrificial layers, the lateral etch rate decreases by < 20 %. The lateral etch rate decrease with increasing etch channel length, which clearly points to a diffusion-limited process as it was already discussed in the literature for (100)-oriented layers.[9,11,14] Probably the etch channel is insufficiently bent open.

A clear trend is observed analyzing the influence of the layer thickness on the lateral etch rate. All measured lateral etch rates of the 50 nm thick $Al_xGa_{1-x}As$ sacrificial layers are larger compared to the 25 nm thick sacrificial layers. We also tried to completely lift off GaAs-(111) layers by etching 10 nm $Al_xGa_{1-x}As$-(111) sacrificial layers, but we observed no lift-off because the lateral etch rate was too low. A thinner etch channel is prominent to be clogged faster by the etch reaction products.[9,11,14] Thus, a thicker sacrificial layer or a more sophisticated method to more severely bend away the released part of the heterostructure [15] should be used for layer lift-off.

The influence of the etch channel width is also confirmed by experiments employing method (1). In addition, the advantage of the wax stressor can be seen: Without wax (method 2), the lateral etch rates drop significantly after a few minutes of etching. E.g., etching of ≤ 25 nm thin $Al_xGa_{1-x}As$-(111) sacrificial layers already stops after roughly 10 min and no lift-off was possible. Without wax, on the one hand, the under-etched layer is not stabilized and might stick to the substrate, and on the other hand, the etch channel is not spread, so the diffusion of reaction products is hampered.[9,11,14]

For the quality of entirely lifted-off membranes it is not relevant, whether method (1) or method (2) was employed. However, for method (1), the lift-off is significantly faster and allows the lift-off of larger membranes. **Figure 5** shows an optical microscopy image of a 500 nm thick GaAs membrane lifted-off and bonded to a glass substrate. The surface is flat and, to a large extent, free of defects or pits. The atomic force microscopy image in Figure 1 b) reveals a smooth surface with a roughness very similar to the surface roughness of the as-grown sample ($\leq 0.5$ nm for $1\times1$ µm$^2$ and $\leq 0.8$ nm for $10\times10$ µm$^2$). The crystal quality of the transferred layer was checked by standard high-resolution x-ray diffraction measurements. Rocking curve measurements reveal full-width half maximum (FWHM) values of 140 arcsec for the bonded film in contrast to 43.2 arcsec for the as-grown film. This shows that the epitaxial lift-off and bonding process barely impairs the structural quality of the film.

To test, whether the transferred films are suitable for nano-pattering, an array of pillars with a pillar diameter of 365 nm and a pitch of 665 nm was patterned into the transferred membrane. Employing a SiN hard mask defined by electron beam lithography, we etched the sample until the GaAs between the pillars were completely removed and the glass became exposed. This can be verified by the contrast in the electron microscope. **Figure 6** shows the electron microscopy image of the pattern. It can be seen that over a large area, the pattern shows no missing pillars. So not only the complete membrane sticks to the glass substrate but also individual pillars. However, one also sees that the pillar diameter is not constant throughout the patterned area. This is not optimal for optical experiments and the etching parameters have to be optimized further.

## 4. Conclusion

We achieved epitaxial $Al_xGa_{1-x}As$-(111) heterostructure growth for $0.5 \leq x \leq 1$ on GaAs-(111)B substrates with smooth surfaces and tested selective etching of Al-containing layers. The lateral etch rate of 50 nm thick AlAs sacrificial layers in 10 % hydrofluoric acid was determined to approximately $(625\pm100)$ µm h$^{-1}$. Whereas for the (111)-oriented layers, the lateral etch rate stays almost constant down to 70 % Al-content before dropping significantly, it drops monotonically for (100)-oriented layers. No etching was observed for aluminum concentrations $\leq 50$ % for both orientations. Besides, we found that a thinner sacrificial layer leads to a lower lateral etch rate. Also, a longer lateral etch depth $d_{lateral}$ results in a reduced lateral etch rate. The wax improves the lift-off process significantly. We were able to lift-off thin GaAs-(111) layers and transfer them to glass substrates. Nano pillars with diameters of 365 nm were fabricated by patterning the transferred film. The array shows on a large area no vacancies, which proves that not only the overall membrane sticks to the surface but each nanometer-scale fraction of the membrane. This highlights the high quality of the lift-off and bonding process presented here. Thus, the presented process is well suitable to fabricate GaAs-(111) nanostructures on glass substrates.

**Acknowledgments**

The work was funded by the Deutsche Forschungsgemeinschaft (DFG, German Research Foundation) - CRC TRR142/2-2020 - Grant No. 231447078 within the sub-projects A06 and C05 and the European Research Council (ERC) under the European Union's Horizon 2020 research and innovation programme (grant agreement No 724306).

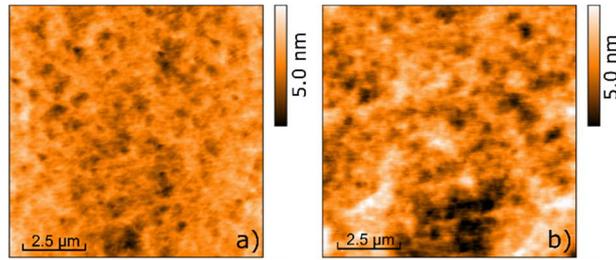

**Figure 1.** Atomic force microscopy images of a) the as-grown GaAs-(111) film and b) the GaAs-(111) film. In both images, the surface roughness (RMS roughness) is very similar (0.6 nm for a) and 0.8 nm for b)).

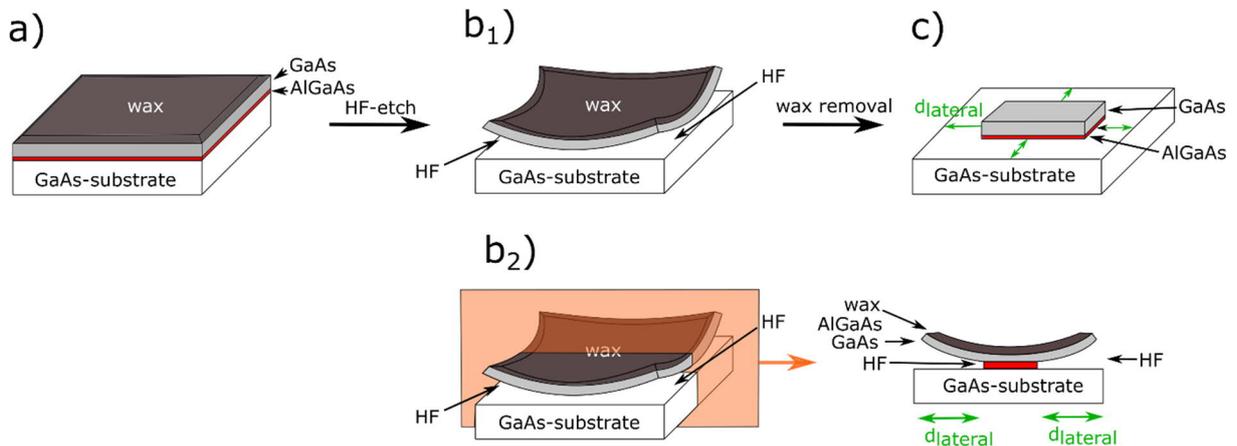

**Figure 2.** Scheme of the sample preparation for determination of lateral $Al_xGa_{1-x}As$ sacrificial layer etch rates. Wax is applied to stabilize the structure and bend the etching channel open. The $Al_xGa_{1-x}As$ sacrificial layer is laterally etched by 10% HF. b2) shows a cross section along the orange plane and the lateral etch depth $d_{lateral}$ is indicated by green arrows. After wax removal in an ultrasonic bath by toluene, the under-etched layer part is removed and only the central not under-etched part remains on the sample (see c)). The lateral etch depth $d_{lateral}$ is determined from the situation shown in c) by optical microscopy.

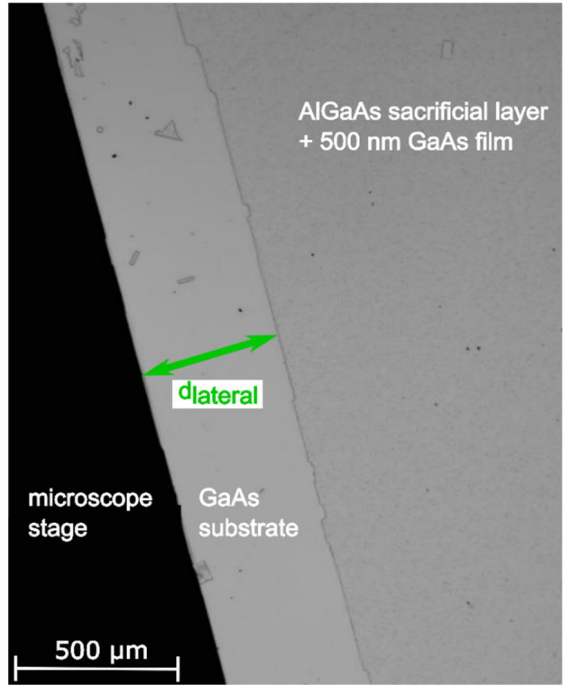

**Figure 3.** Optical microscopy image (top view) showing the edge region of a sample prepared by method (1) (see text). The sample status shown corresponds to the situation depicted in Figure 2c. The green arrow indicates the lateral etch depth as also defined in Figure 2.

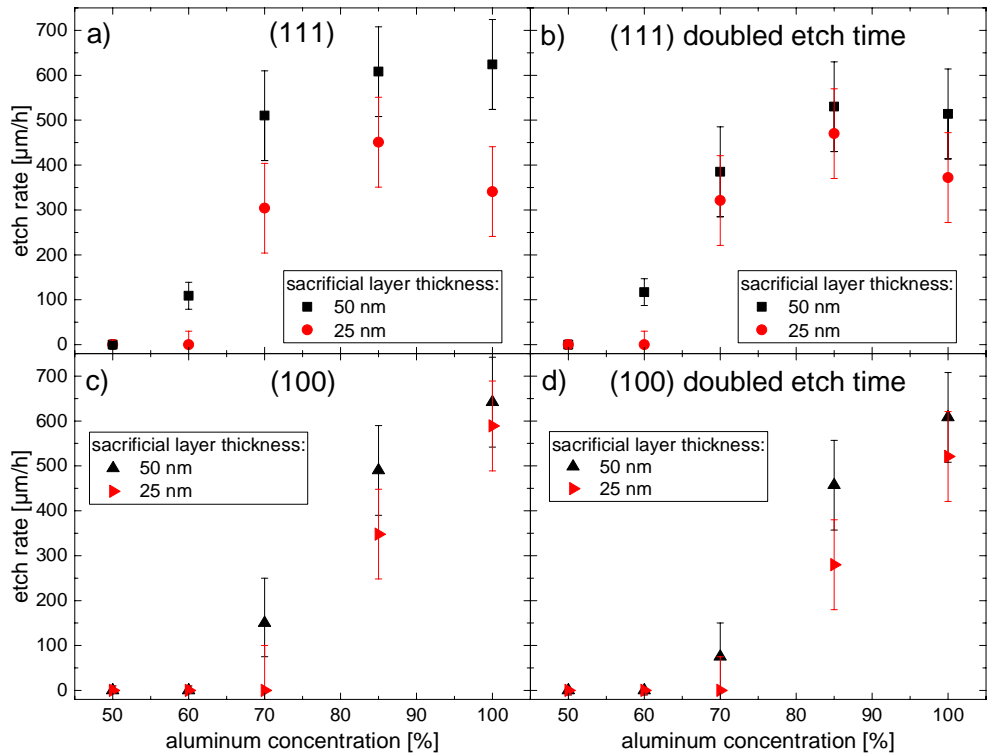

**Figure 4.** Lateral etch rates for (111)- and (100)-oriented $Al_xGa_{1-x}As$ sacrificial layers as a function of the Al-content. The etch rates have been determined with method (1) for 25 nm and 50 nm thick sacrificial layers. The etch times were for most Al concentrations 65 (4a and 4c) and 130 (4b and 4d) minutes, respectively. Only for $x \leq 60\%$ longer etching times up to 260 minutes have been employed.

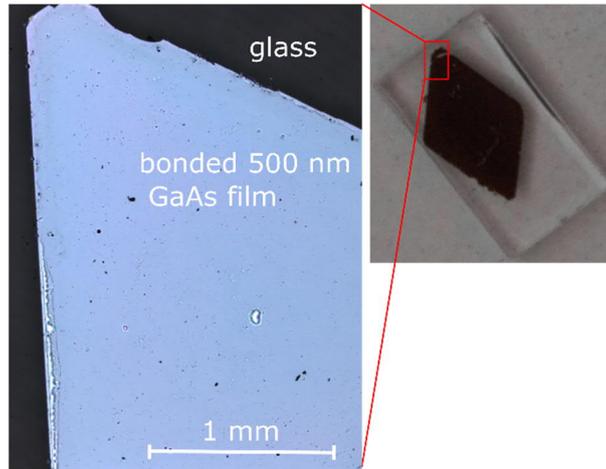

**Figure 5.** The figure shows a 500 nm thick GaAs-(111)B layer lifted-off and transferred to a glass substrate. Right: Overview image of the whole transferred film (4×4 mm$^2$) and the supporting glass substrate. Left: Optical microscopy image (top view) showing a part (indicated by a red box in the right image) of the sample with higher magnification.

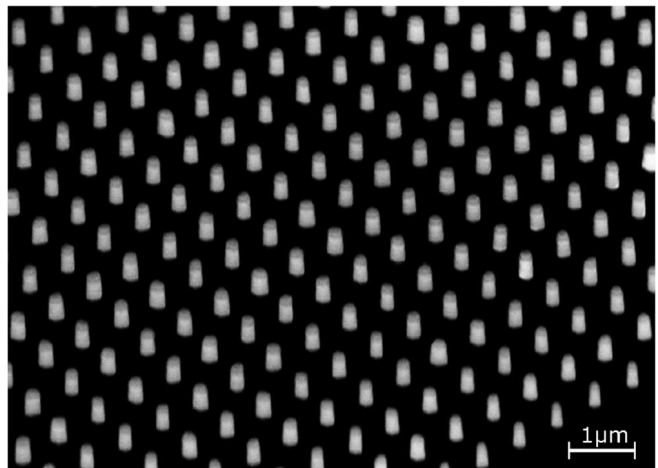

**Figure 6.** Scanning electron microscopy image (under 45 °) of GaAs-(111)B pillars on transparent glass. The GaAs-(111)B film was lifted off the substrate and bonded to the glass substrate before patterning. The slightly darker area on top of each pillar is Silicon nitride, which we used as a hard mask to transfer the geometry to the GaAs-(111)B thin film. To image the pillars on the insulating glass, the whole sample was covered by a 6 nm thick carbon film.